\documentclass[aps,prd,twocolumn]{revtex4-2}
\usepackage{xpatch}
\makeatletter
\xpatchcmd{\@ssect@ltx}{\@xsect}{\protected@edef\@currentlabelname{#8}\@xsect}{}{}
\xpatchcmd{\@sect@ltx}{\@xsect}{\protected@edef\@currentlabelname{#8}\@xsect}{}{}
\makeatother
\usepackage{hyperref}
\usepackage{amsmath}
\usepackage{amssymb}
\usepackage{graphicx}
\usepackage{xcolor}

\begin{document}

\title{From square plaquettes to triamond lattices for SU(2) gauge theory}

\author{Ali H. Z. Kavaki}\email{alihzk@yorku.ca}
\author{Randy Lewis}\email{randy.lewis@yorku.ca}
\affiliation{Department of Physics and Astronomy, York University, Toronto, ON M3J 1P3, Canada}

\date{June 28, 2024}

\begin{abstract}
Lattice gauge theory should be able to address significant new scientific questions when implemented on quantum computers.
In practice, error-mitigation techniques have already allowed encouraging progress on small lattices.
In this work we focus on a truncated version of SU(2) gauge theory, which is a familiar non-Abelian step toward quantum chromodynamics.
First, we demonstrate effective error mitigation for imaginary time evolution on a lattice having two square plaquettes,
obtaining the ground state using an IBM quantum computer and observing that this would have been impossible without error mitigation.
Then we propose the triamond lattice as an expedient approach to lattice gauge theories in three spatial dimensions and we derive
the Hamiltonian.
Finally, error-mitigated imaginary time evolution is applied to the three-dimensional triamond unit cell, and its
ground state is obtained from an IBM quantum computer.
Future work will want to relax the truncation on the gauge fields, and the triamond lattice is increasingly valuable for such studies.
\end{abstract}

\maketitle


Quantum field theory combines quantum mechanics with special relativity and is of
widespread importance in high-energy physics, nuclear physics, and condensed matter physics.
Enforcing a spatially local symmetry leads to the special case of a gauge theory.
The standard model of particle physics is a collection of three gauge theories describing the strong, weak and
electromagnetic forces.
Because perturbation theory cannot be applied to a large coupling constant, many aspects of the strong force are
best described by a numerical approach called lattice gauge theory.

Lattice gauge theory has been of central importance to quantum chromodynamics for several decades and has become a precision tool for
practitioners \cite{Gattringer:2010zz,Knechtli:2017sna}.
It provides rigorous calculations of static properties, such as the masses and form factors of hadrons, directly from
the gauge theory of quarks and gluons.
Such calculations are crucial in the ongoing quest to understand hadron physics, such as the tetraquarks and pentaquarks
that have been discovered experimentally in recent years \cite{Brambilla:2019esw,Bicudo:2022cqi}.
Lattice gauge theory calculations also provide necessary input for experiments seeking new physics beyond the standard model,
where a famous recent example is the anomalous magnetic moment of the muon \cite{Aoyama:2020ynm}.

Quantum computers offer a prospective way to calculate more than just static properties.
Traditional lattice gauge theory uses Monte Carlo methods that would have sign problems if applied straightforwardly to situations involving dynamics, nonzero density, or
some topological interactions \cite{Nagata:2021ugx,Funcke:2023jbq}.
These sign problems can be avoided by using a Hamiltonian approach, where the exponentially large Hilbert space can fit naturally onto a quantum computer.
Now that quantum computing hardware is becoming a reality, this powerful new approach to lattice gauge theory is being developed by a broad community of researchers
\cite{Funcke:2023jbq,Bauer:2022hpo,DiMeglio:2023nsa}.

With quantum chromodynamics as a motivating long-range goal, the present work is focused on the simplest non-Abelian gauge theory.
Being non-Abelian means that gauge fields carry the relevant charge directly and can therefore interact among themselves, like the gluons of quantum chromodynamics
and in contrast to the photons of quantum electrodynamics.
Specifically our work will consider SU(2) gauge theory in the absence of fermions, and we will truncate the gauge field to fit on a small number of qubits.
Several research groups have already carried out exploratory studies of non-Abelian gauge theories on quantum computing hardware
\cite{Banerjee:2012xg,Klco:2019evd,Ciavarella:2021nmj,Atas:2021ext,ARahman:2021ktn,Ciavarella:2021lel,Illa:2022jqb,ARahman:2022tkr,Farrell:2022wyt,Atas:2022dqm,Farrell:2022vyh,Ciavarella:2023mfc}.
There have also been many theoretical studies laying the groundwork for anticipated implementations on larger quantum computers
\cite{Kogut:1974ag,Chandrasekharan:1996ih,Mathur:2004kr,Byrnes:2005qx,Tagliacozzo:2012df,Zohar:2012xf,Zohar:2013zla,Stannigel:2013zka,Anishetty:2014tta,Zohar:2014qma,Mezzacapo:2015bra,Silvi:2016cas,Banuls:2017ena,Banerjee:2017tjn,Raychowdhury:2018tfj,Sala:2018dui,Raychowdhury:2018osk,Zohar:2019ygc,Raychowdhury:2019iki,Ji:2020kjk,Kasper:2020akk,Davoudi:2020yln,Dasgupta:2020itb,Kasper:2020owz,Kan:2021nyu,Zohar:2021nyc,Halimeh:2021vzf,Raychowdhury:2021jbo,Klco:2021lap,Mildenberger:2022jqr,Gonzalez-Cuadra:2022hxt,Carena:2022hpz,Davoudi:2022xmb,Yao:2023pht,Jakobs:2023lpp,Zache:2023dko,Hayata:2023puo,Halimeh:2023wrx,Chan:2023fwv,Bauer:2023qgm,Muller:2023nnk,Cataldi:2023xki,Bauer:2023jvw,Ebner:2023ixq,Ebner:2024mee,Gustafson:2023kvd,Turro:2024pxu,Illa:2024kmf}.
In the present work, we address two important issues: an effective way to mitigate errors and a practical way to extend lattices into three spatial dimensions.

Future quantum computers are expected to have smaller error rates and robust methods for correcting those errors, but calculations in the present era
face substantial error rates and typically have insufficient resources to implement true error correction.
Instead, error mitigation methods have been devised \cite{Endo:2020kro}, and these can provide significant improvements for computations performed on today's hardware.
Our study demonstrates the use of an existing method called self-mitigation \cite{ARahman:2022tkr} but in a new context: evolution of a quantum state through imaginary time.
As is well known, after sufficiently many steps of imaginary time the excited state contributions to a generic initial state will become exponentially suppressed
relative to the ground state, thus providing a way to create the ground state and to determine its eigenvalue.
Our results will demonstrate that self-mitigation correctly finds and sustains the true ground state of a two-plaquette lattice
even though the unmitigated data points are moving ever farther from the true result with each additional time step.

With successful mitigation in hand, we then turn our attention to lattice gauge theory in three spatial dimensions.
While this can be done with a standard cubic lattice, where six gauge links touch each lattice site,
there is an important issue to address. Specifically,
for any lattice where more than three gauge links meet at a lattice site,
the quantum numbers of the gauge links themselves are insufficient to fully define the state of the gauge field across the lattice
\cite{Raychowdhury:2018tfj,Raychowdhury:2019iki,Zache:2023dko,Muller:2023nnk}.
To understand why, consider a two-dimensional square lattice where four SU(2) gauge links meet at every site.
At one particular site, suppose each of the four gauge links happen to have
quantum number $j=\frac{1}{2}$, where $j$ is notation familiar from quantized angular momentum
though here it refers to the gauge degrees of freedom.
One pair of these links could sum to $j_{\rm pair}=0$ or $1$.
The other pair must have the same sum because Gauss's law requires the total of all four to be $j_{\rm site}=0$.
Therefore a full description of the state requires not just the $j$ values of individual links, but also
the value of $j_{\rm pair}$ for some pair selected by the user to define that site.

One way to handle this issue is to assign a sufficient number of additional qubits at each lattice site, being careful to define a convention at each site, thereby
completing the state's definition \cite{Raychowdhury:2018tfj,Raychowdhury:2019iki,Zache:2023dko,Muller:2023nnk}.
The number of additional qubits needed to define $j_{\rm pair}$ grows logarithmically
with the user’s choice for $j_{\rm max}$.  Moreover, a three-dimensional lattice
needs more than a single $j_{\rm pair}$ at each site.
In the present work we propose an alternative based on a structure from crystallography \cite{Laves:1932,Coxeter:1955,Sunada:2008,Sequin:2008,Conway:2008,Suizu:2022}.
Having only three links at each site, our choice will avoid the need for any quantum numbers beyond the individual link
values.

Consider a lattice where each site is touched by exactly three gauge links that have equal lengths, lie in a single plane,
and are placed at equal angles around the site, i.e.\ $2\pi/3$.
This forms a three-dimensional lattice that has a high degree of symmetry and needs no additional qubits beyond the links themselves.
Because of its similarity to a diamond crystal of carbon atoms but with three links per site instead of more,
it has been called the triamond lattice \cite{Sequin:2008,Conway:2008}.
It is also known as the Laves lattice \cite{Coxeter:1955} or the $K_4$ lattice \cite{Sunada:2008,Suizu:2022}.
In this work, we provide an introduction to lattice gauge theory on the triamond framework.
We derive the SU(2) Hamiltonian for a triamond lattice and demonstrate its use on a noiseless simulator.
Then we employ our method for error-mitigated imaginary time
evolution to create the ground state of the three-dimensional
triamond unit cell on an IBM quantum computer \cite{ibm}.

\section*{Results}\label{sec:results}

\subsection*{Imaginary time evolution on square plaquettes}

Finding the ground state for a given Hamiltonian is an important ingredient of many scientific studies.
Two common approaches are variational methods and imaginary time evolution.
Variational methods rely on an ansatz chosen by the user, and can only get as close to the true ground state as the ansatz will allow.
In contrast, imaginary time evolution always converges to the true ground state unless the initial state is perfectly orthogonal to it.

Imaginary time evolution is used routinely in traditional lattice gauge theory on classical computers.
Our motivation to study it now on a quantum computer is not to compete with successful classical methods, but rather
to imagine that ground-state preparation will become the first step in a larger computation that truly does require quantum computing
\cite{Chai:2023qpq,Farrell:2024fit,Davoudi:2024wyv}.
Imaginary time evolution has recently been applied to $Z_2$ gauge theory as well \cite{Davoudi:2022uzo}.

\begin{figure}
\includegraphics[width=85mm]{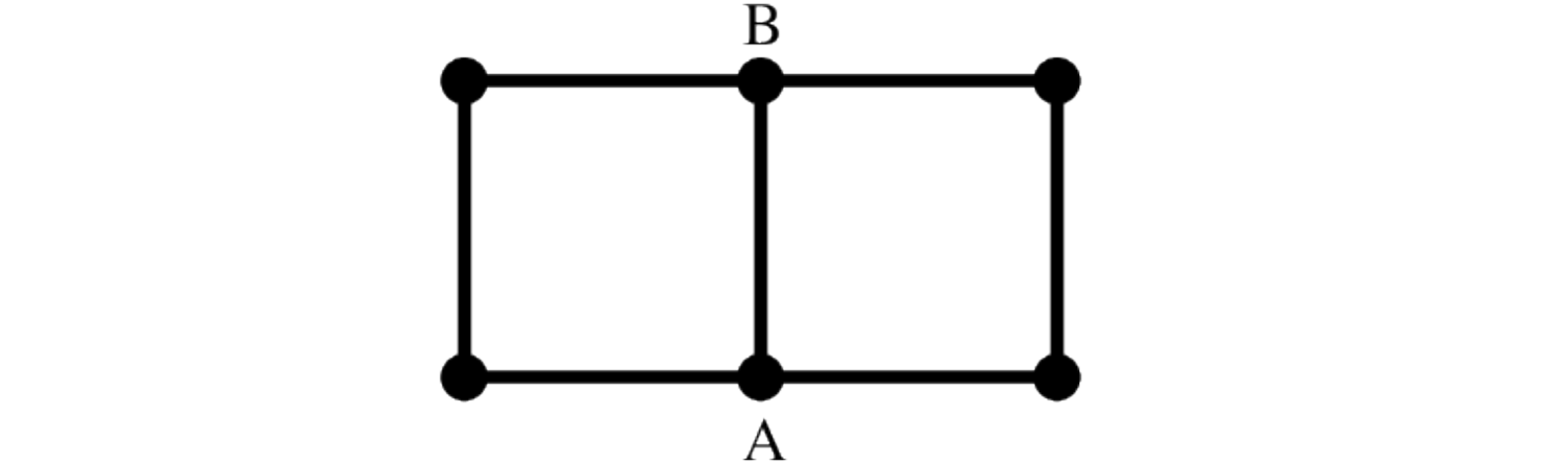}
\caption{{\bf A minimal lattice.} This is the first lattice used to demonstrate self-mitigation for quantum imaginary time evolution.
         The text refers to the 3 paths between sites A and B as the left path (three links), the center path (one link),
         and the right path (three links).\label{fig:squarepair}}
\end{figure}

Our first computation uses a lattice comprising two side-by-side square plaquettes that share a single gauge link, thus forming
a left path, a center path, and a right path as shown in Fig.~\ref{fig:squarepair}.
Each of the seven gauge links on this lattice is a superposition of SU(2) representations, and Gauss's law requires the three on the left path
to equal one another and similarly the three on the right path must equal one another.
For our calculation we truncate the basis and retain only the two lowest states for each gauge link, $j=0$ and $j=\frac{1}{2}$.

Any state of this system can be described by two qubits, one for the left path and one for the right path, because Gauss's law then determines
the center path uniquely.
Such a tiny system is a valuable place to study error mitigation because a single time step requires only a simple circuit while more and more time steps will
require more and more entangling gates.
The real-time evolution of this system was studied previously as the first example of self-mitigation \cite{ARahman:2022tkr}.
The Hamiltonian, in units of $g^2/2$, is
\begin{eqnarray}
\hat H &=& \frac{3}{8}\left(7-3\hat Z_0-\hat Z_0\hat Z_1-3\hat Z_1\right) - \frac{x}{2}\left(3+\hat Z_1\right)\hat X_0 \nonumber \\
       & & - \frac{x}{2}\left(3+\hat Z_0\right)\hat X_1 \,, \label{eq:H}
\end{eqnarray}
where $x=2/g^4$ and $g$ is the coupling constant, i.e.\ the strength of the strong interaction for this physics theory.
The Pauli gates acting on qubit $i$ are $\hat X_i$, $\hat Y_i$ and $\hat Z_i$.

Imaginary time evolution is not a unitary operation, so how can it be implemented on a quantum computer?
We use an algorithm developed by Motta et al., called quantum imaginary time evolution (QITE) \cite{Motta:2019yya}, which can be described as follows.
The evolution we seek is
\begin{equation}
\left|\Psi(\tau)\right> = e^{-\tau\hat H}\left|\Psi(0)\right> \,, \label{eq:Hstep}
\end{equation}
where $\tau$ is the real-valued Euclidean time parameter.
We can define a normalized state $\left|\psi\right>$ such that $\left|\Psi(0)\right> = r\left|\psi\right>$ for positive $r$.
The evolution can always be expressed as
\begin{equation}
\left|\Psi(\tau)\right> = r^\prime e^{-i\tau\hat A}\left|\psi\right> \,, \label{eq:Astep}
\end{equation}
for some unitary operator $e^{-i\tau\hat A}$ and positive $r^\prime$.
An immediate consequence is
\begin{equation}\label{eq:rprime}
r^\prime = r\left(1-\tau\left<\psi\right|\hat H\left|\psi\right>\right) + O(\tau^2) \,.
\end{equation}
The operator $\hat A$ can be determined from a state tomography of $\left|\psi\right>$ that reflects the specific qubit connectivities of $\hat H$.

For our physics example, Eq.~(\ref{eq:H}) confirms that the Hamiltonian is purely real, meaning that we can restrict ourselves to real-valued basis states as well.
Together with Eqs.~(\ref{eq:Hstep}) and (\ref{eq:Astep}), this realness means $\hat A$ must be purely imaginary, leading to a general expression with odd powers of $\hat Y_i$ gates,
\begin{eqnarray}
\hat A &=& a_{iy}\hat Y_0 + a_{xy}\hat X_1\hat Y_0 + a_{zy}\hat Z_1\hat Y_0 \nonumber \\
       & & + a_{yi}\hat Y_1 + a_{yx}\hat Y_1\hat X_0 + a_{yz}\hat Y_1\hat Z_0\,. \label{eq:Adef}
\end{eqnarray}
The state tomography that determines the values of the coefficients $a_{jk}$ is presented in \nameref{sec:methods}.
With those values in hand, and with $r^\prime$ obtained from the calculation of Eq.~(\ref{eq:rprime}),
the state at a small value of $\tau$ can be computed from Eq.~(\ref{eq:Astep}).
In particular, we use
\begin{eqnarray}
e^{-i\theta\hat Y_j} &=& RY_j(2\theta) \,, \\
e^{-i\theta\hat X_j\hat Y_k} &=& CX_{kj}RY_k(2\theta)CX_{kj} \,, \\
e^{-i\theta\hat Z_j\hat Y_k} &=& CX_{jk}RY_k(2\theta)CX_{jk} \,,
\end{eqnarray}
where $CX_{jk}$ is a controlled not (CNOT) gate with qubit $j$ as control and qubit $k$ as target.

\begin{figure}
\includegraphics[width=85mm]{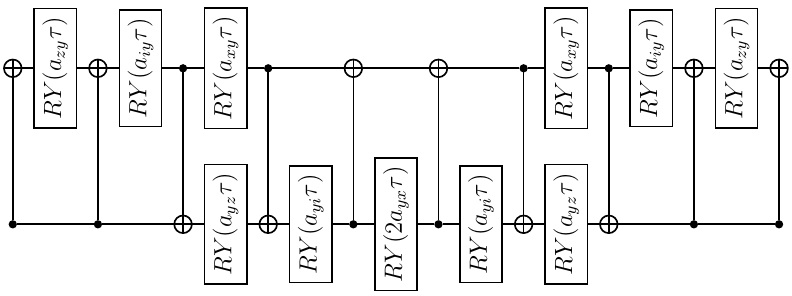}
\caption{{\bf One Trotter step.} This circuit is used for one second-order Trotter step of imaginary time evolution for SU(2) gauge theory on a lattice comprising two square plaquettes.
$\tau$ is the real-valued Euclidean time parameter. The coefficients $a_{jk}$ define time evolution as shown in Eq.~(\ref{eq:Adef}).
         \label{fig:circuit}}
\end{figure}

For sufficiently small $\tau$, the operator $e^{-i\theta\hat A}$ can be approximated by a product of six factors, one for each term in $\hat A$.
For a better approximation, that first-order Trotter expression can be replaced by the second-order Trotter expression represented by the circuit
shown in Fig.~\ref{fig:circuit}.
To reach a larger value of $\tau$, we can perform a sequence of small steps where the circuit is several end-to-end copies of Fig.~\ref{fig:circuit}.
Notice that the tomography needs to be computed separately for each step, using the state at the previous step as input.

\begin{figure}
\includegraphics[width=85mm]{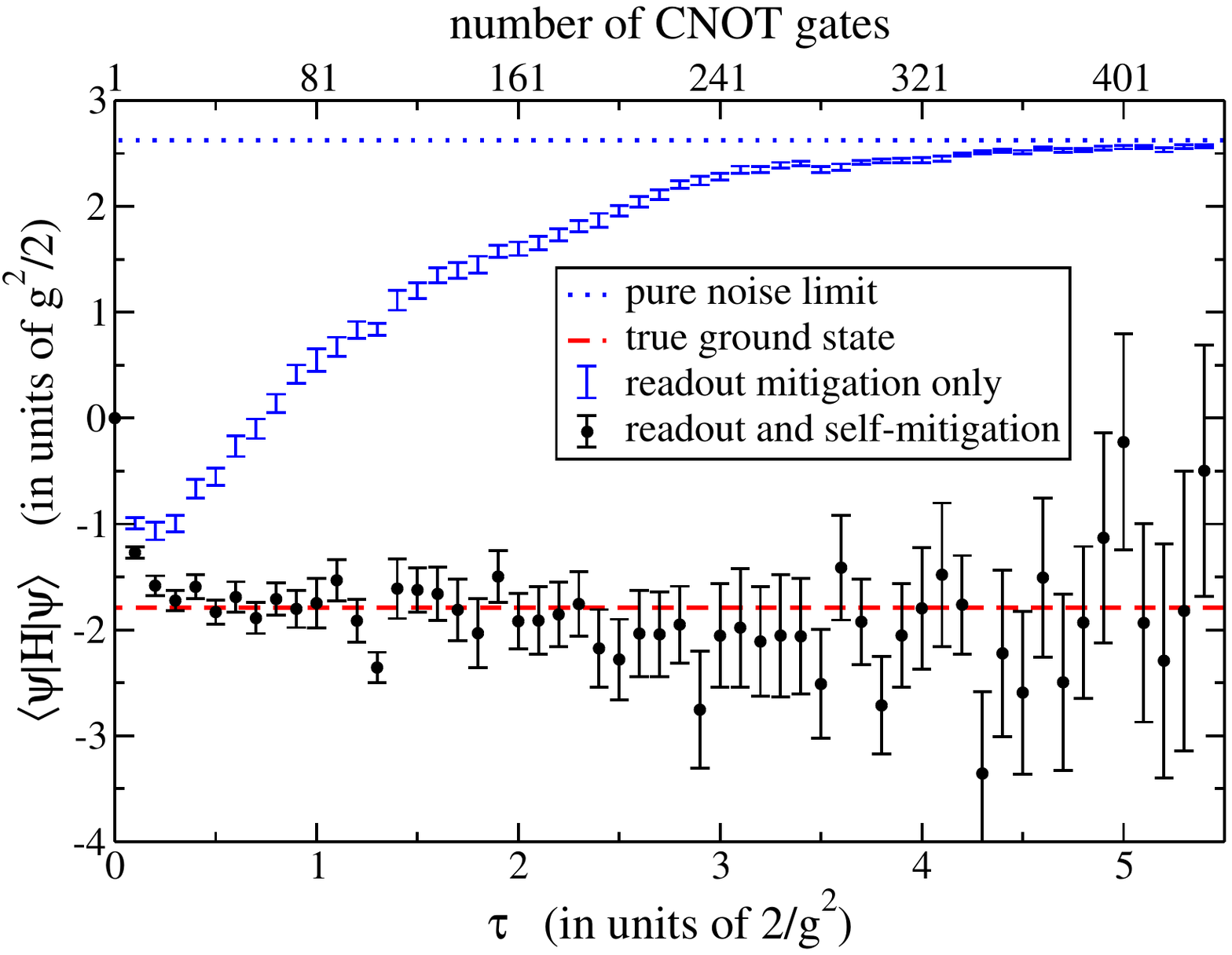}
\caption{{\bf Quantum imaginary time evolution with and without self-mitigation.}
         These results are for SU(2) gauge theory with $j_{\rm max}=\frac{1}{2}$ on a lattice
         comprising two square plaquettes.
         The gauge coupling is $x=1$ and the time step is $\Delta\tau=0.1$ in units of $2/g^2$.
         Each quantum circuit uses 50 randomized compilings with $10^4$ shots per compiling.
         All quantum computations were performed on {\tt ibm\_lagos}.
         Black data points use self-mitigation and blue data points do not.
         Error bars are 95\% confidence intervals.
         The true ground state is shown as a red dashed line.
         The pure noise limit is shown as a blue dotted line.
         \label{fig:plot2mitb}}
\end{figure}

Because the two-qubit CNOT gate is noisier than a single-qubit gate on quantum hardware, the gates in Fig.~\ref{fig:circuit} have been ordered
in a way that minimizes the number of CNOT gates.
Also, the CNOT gate at the end of one Trotter step will cancel the CNOT gate at the beginning of the next step.
As well, the very first CNOT gate in the circuit can be omitted because the initial control qubit is always off.
Nevertheless, the error bars without symbols in Fig.~\ref{fig:plot2mitb} show that only the first two or three time steps approach the true ground state energy and then subsequent
time steps move further and further away.
The quantum computation is overwhelmed by hardware noise without ever reaching the correct result.

Self-mitigation is able to extract the true result from the computed data.
The original implementation \cite{ARahman:2022tkr}, which was inspired by the work of Ref.~\cite{Urbanek:2021oej}, has subsequently been used and extended in various ways
\cite{Farrell:2022wyt,Atas:2022dqm,Ciavarella:2023mfc,Chan:2023fwv,Charles:2023zbl,Farrell:2023fgd,Asaduzzaman:2023wtd,Hidalgo:2023wzr,Farrell:2024fit,Kiss:2024foh}.
The basic idea is to run a mitigation circuit that is identical to the physics circuit except that $\tau\to-\tau$ in the second half of the circuit.
If your circuit has an odd number of second-order
Trotter steps, use $2\tau\to0$ in the center gate (see, for example, Fig.~\ref{fig:circuit})
and insert a barrier to prevent CNOT cancellation.

The true outcome of the mitigation circuit should be identical to the initial state, i.e.\ all qubits should be in the off position.
Comparison of the computed results with this known true result provides a measurement of hardware errors.
The physics circuit will have similar hardware errors because the two circuits are identical up to the sign of $\tau$ in the latter half.

More precisely, randomized compiling (see \nameref{sec:methods}) is used to convert CNOT errors into incoherent noise that is
well described by the depolarizing noise model \cite{Urbanek:2021oej}.
For our example of imaginary time evolution, this leads to a pair of ratios,
\begin{equation}
\left.\frac{\left<\hat P_j\right>_{\rm true}}{\left<\hat P_j\right>_{\rm comp}}\right|_{\rm physics~run} =
\left.\frac{1}{\left<\hat Z_j\right>_{\rm comp}}\right|_{\rm mitigation~run}, \label{eq:ratio1}
\end{equation}
\begin{equation}
\left.\frac{\left<\hat P_j\hat P_j\right>_{\rm true}}{\left<\hat P_j\hat P_k\right>_{\rm comp}}\right|_{\rm physics~run} =
\left.\frac{1}{\left<\hat Z_j\hat Z_k\right>_{\rm comp}}\right|_{\rm mitigation~run}, \label{eq:ratio2}
\end{equation}
where $\hat P_j\in\{\hat X_j,\hat Y_j,\hat Z_j\}$.
These equations give the true expectation value of a Pauli string as the ratio of results computed by a physics run and a mitigation run.
They are the counterparts of Eq.~(8) in the original paper \cite{ARahman:2022tkr} but look slightly different because the previous work studied probabilities that equalled
$\frac{1}{2}$ in the pure noise limit whereas the present work employs expectation values of Pauli strings that are zero for pure noise.

The filled data points of Fig.~\ref{fig:plot2mitb} were obtained by computing each term of Eq.~(\ref{eq:H}) from the ratios in Eqs.~(\ref{eq:ratio1})
and (\ref{eq:ratio2}).
The physics runs and mitigation runs (with 50 randomized compilings for each) and the readout error mitigation runs were all completed within a single {\tt qiskit}
job \cite{qiskit} to ensure they were experiencing the same hardware conditions.
Whereas the unmitigated data in Fig.~\ref{fig:plot2mitb}
approach the pure noise limit of $\frac{21}{8}$, where only the first term in Eq.~(\ref{eq:H})
makes a nonzero contribution,
the self-mitigated data points in Fig.~\ref{fig:plot2mitb} remain consistent with the true ground state as $\tau$ increases.
Error bars grow as the circuit gets longer, but meaningful results are still obtained with more than 400 CNOT gates.
This success provides an incentive to move toward larger lattices, including novel three-dimensional approaches like
the triamond lattice.

\subsection*{The SU(2) Hamiltonian for a triamond lattice}\label{sec:triH}

\begin{figure}
\includegraphics[width=88mm]{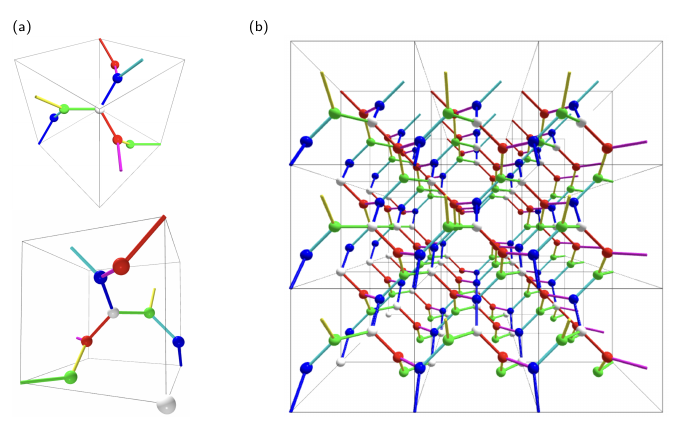}
\caption{{\bf The triamond lattice.} Panel (a) shows two views of the unit cell, which always contains 8 sites and 12 gauge links.
         Panel (b) shows a lattice having 3 unit cells along each direction.
         In both panels, the colour of each link defines its direction according to Eqs.~(\ref{eq:colorvectors1}) and (\ref{eq:colorvectors2}).
         \label{fig:tripanels}}
\end{figure}
Fig.~\ref{fig:tripanels} shows the locations of lattice sites and gauge links on a triamond lattice.
See also our two-minute video presentation (\url{https://vimeo.com/904585432}), which provides views of the triamond lattice from a camera moving around it.

To introduce the triamond lattice quantitatively, we will first define six
unit vectors labeled as colors (red, green, blue, cyan, magenta, yellow) to aid the discussion,
\begin{align}
\hat r &= \frac{\hat j - \hat k}{\sqrt{2}}, &
\hat g &= \frac{\hat i - \hat j}{\sqrt{2}}, &
\hat b &= \frac{\hat k - \hat i}{\sqrt{2}}, \label{eq:colorvectors1}
\end{align}
\begin{align}
\hat c &= \frac{\hat j + \hat k}{\sqrt{2}}, &
\hat m &= \frac{\hat i + \hat j}{\sqrt{2}}, &
\hat y &= \frac{\hat k + \hat i}{\sqrt{2}}, \label{eq:colorvectors2}
\end{align}
where $\hat i$, $\hat j$ and $\hat k$ are the standard orthonormal vectors.
Every gauge link in the triamond lattice is along one of these color directions.
As an aside, we note that the colors defined here have nothing to do with the colors of quantum chromodynamics.

Each lattice site is touched by three gauge links that lie in a plane.
The complete lattice has just four sets of parallel planes, so we color each site white, red, green or blue to represent which plane is at that site.
The links and sites form a color scheme that is motivated by the well-known RGB model of colors: each white site is touched by red, green and blue links, each red site is touched by red, magenta and yellow links, each green site is touched by green, cyan and yellow links, and each blue site is touched by blue, cyan and magenta links.

The angle between any two planes is $\arccos(\frac{1}{3})$, which is familiar because this same angle lies between any pair of lines joining the center of a cube to its corners.
Because the unit vector orthogonal to a plane is only unique up to its sign, the triamond lattice is chiral.
Specifically, the mirror image of Fig.~\ref{fig:tripanels} is a distinct but equally valid triamond lattice.

Notice that the white sites of a triamond lattice form a body-centered cubic (bcc) structure.
It is useful to compare the options of defining a lattice gauge theory on a triamond lattice, on a bcc lattice, or on a simple cubic lattice.
Relative to a simple cubic lattice of the same volume, a bcc lattice has twice as many lattice sites, gauge links that are shorter by a factor of $\sqrt{2}$, and each bcc site has 8 nearest neighbors compared to 6 for the simple cubic lattice.
This suggests that a bcc lattice could be valuable for efficient discussions of the continuum limit, perhaps alongside simple cubic studies.
Importantly, the bcc and simple cubic lattices both have more than three gauge links touching each lattice site, meaning that extra qubits would be needed within each lattice site (and a convention must be chosen) to fully define the quantum state of the lattice \cite{Raychowdhury:2018tfj,Raychowdhury:2019iki,Zache:2023dko}.
In a sense, the triamond lattice can be viewed as a bcc lattice where those implicit qubits have been made explicit and unambiguous
by creating the red, green and blue sites.

Although the white sites do form a bcc lattice,
the triamond lattice is more economical than merely being a clarified bcc lattice.
Imagine, for example, shrinking the red, green and blue links into their white sites.
The remaining lattice would not be bcc because it would have only 6 links per unit cell (2 cyan, 2 magenta, and 2 yellow)
whereas a bcc lattice would require 8 links per unit cell.

The comparison to bcc is encouraging, but our primary motivation for proposing the triamond lattice is
the fact that it has exactly three links touching each site, so the $j$ values of each gauge link are sufficient to define
basis states for the lattice.
For planar physics, a hexagonal lattice succeeds in having three links touching each site, and SU(2) gauge theory
has recently been implemented on a hexagonal lattice \cite{Muller:2023nnk,Ebner:2023ixq}.
Notice that the shortest closed paths on a hexagonal lattice have six gauge links.

The shortest closed paths on a triamond lattice have 10 gauge links. We refer to these as elementary plaquettes.
Among the 10 gauge links comprising any elementary plaquette, one of the 6 colors is absent and the other 5 colors always occur twice.
The lattice has only 6 orientations of plaquettes (the non-red plaquette, the non-green plaquette, etc.), with copies of
them translated spatially throughout the lattice.

To perform any quantum computation on a triamond lattice, we need a Hamiltonian.
For SU(2) gauge theory, the Hamiltonian is the non-Abelian generalization of quantum electrodynamics and is written
as a sum of electric terms and magnetic terms \cite{Kogut:1974ag}.
In the continuum limit, the Hamiltonian is
\begin{eqnarray}
H^{\rm cont} &=& H_E^{\rm cont} + H_B^{\rm cont} \,, \\
H_E^{\rm cont} &=& g^2\int{\rm Tr}\left(E_x^2+E_y^2+E_z^2\right)d^3x \,, \label{eq:HEcont} \\
H_B^{\rm cont} &=& \frac{1}{g^2}\int{\rm Tr}\left(F_{xy}^2+F_{yz}^2+F_{zx}^2\right)d^3x \,, \label{eq:HBcont}
\end{eqnarray}
where each trace is over the SU(2) indices of the electric field $\vec E$ or field strength tensor $F_{\mu\nu}$.
(Other normalization conventions are sometimes used for $\vec E$ and $F_{\mu\nu}$ \cite{Muller:2023nnk}.)

On a lattice, each gauge link is an element of SU(2),
\begin{equation}\label{eq:U}
U(\vec n,\hat s) = e^{ia\hat s\cdot\vec A(\vec n)} \,,
\end{equation}
where $a$ is the lattice spacing.
The component of the gauge field at site $\vec n$ that points in direction $\hat s$ is an element of the SU(2) algebra,
\begin{equation}
A_s(\vec n) = \sum_{k=1}^3\sigma^kA_s^k(\vec n) \,.
\end{equation}
Our computations will be performed in the electric basis, where the magnetic terms are best expressed as
a sum of all elementary plaquettes in this lattice
and each plaquette is the trace of the ordered product of the 10 gauge links comprising that plaquette.
The somewhat tedious derivation of this lattice Hamiltonian is outlined in \nameref{sec:methods}.
The result is
\begin{eqnarray}
H &=& H_E + H_B \,, \label{eq:Htri} \\
H_E &=& \frac{8\sqrt{2}a^3g^2}{3}\!\!\!\sum_{n={\rm links}}{\rm Tr}\left(E_x^2(n)+E_y^2(n)+E_z^2(n)\right),~~~~~ \label{eq:HE} \\
H_B &=& -\frac{2\sqrt{2}}{g^2a}\sum_{\vec w={\rm white}}\sum_{s=1}^6\mathcal{P}_s(\vec w) \,, \label{eq:HB}
\end{eqnarray}
where the lattice spacing $a$ is defined to be the distance between nearest-neighbor sites, i.e.\ the length of each
gauge link on the lattice.
The sum in $H_B$ over 6 plaquettes, $\mathcal{P}_s(\vec w)$, for each white lattice site includes all plaquettes on the lattice.
Specifically, there is one of each type of plaquette (non-red, non-green, etc.) associated with each white site, and we
use the subscript $s$ to identify the color that is absent from the plaquette.

Any eigenstate of $H_E$ is fully defined by providing the quantum numbers for every gauge link on the triamond lattice,
$j_1$, $j_2$, $j_3$, \ldots, $j_N$.  Our shorthand for an eigenstate is $\left|\{j\}\right>$.
The diagonal matrix elements of the Hamiltonian are
\begin{equation}\label{eq:ondiag}
\left<\{j\}\right|H_E\left|\{j\}\right> = \frac{8\sqrt{2}g^2}{3a}\sum_{n=1}^Nj_n(j_n+1) \,,
\end{equation}
and the off-diagonal matrix elements from initial state $\left|\{j\}\right>$ to final state $\left|\{J\}\right>$ come from
Eq.~(\ref{eq:HB}) with
\begin{eqnarray}
\left<\{J\}\right|\mathcal{P}_s(n)\left|\{j\}\right> &=& \prod_{\rm sites}(-1)^{j_e+j_f+\frac{1}{2}+J_b}\sqrt{2J_f+1} \nonumber \\
     && \sqrt{2j_b+1}\left\{\begin{array}{ccc} j_e & j_f & j_b \\ \frac{1}{2} & J_b & J_f \end{array}\right\} \,,
        \label{eq:Pelement}
\end{eqnarray}
where the right-hand side has a standard $6j$ symbol and the product includes the 10 sites in order around the plaquette.
For SU(2), either direction around the plaquette gives the same result.
There are three links at each site, one external to the plaquette (subscript $e$),
one in the forward direction around the plaquette (subscript $f$),
and one in the backward direction around the plaquette (subscript $b$).
Eq.~(\ref{eq:Pelement}) is not specific to the triamond lattice and has been obtained previously by other authors
\cite{Zache:2023dko,Hayata:2023puo}.

\subsection*{Computations on the triamond unit cell}

As shown in Fig.~\ref{fig:tripanels}, a unit cell contains 12 gauge links.
By using periodic boundary conditions, a single unit cell becomes a viable three-dimensional lattice having exactly
three gauge links touching each site.
If each SU(2) gauge link is truncated to just two basis states, $j=0$ and $j=\frac{1}{2}$, the unit cell
accommodates $2^{12}=4096$ basis states.

\begin{figure}
\includegraphics[width=85mm]{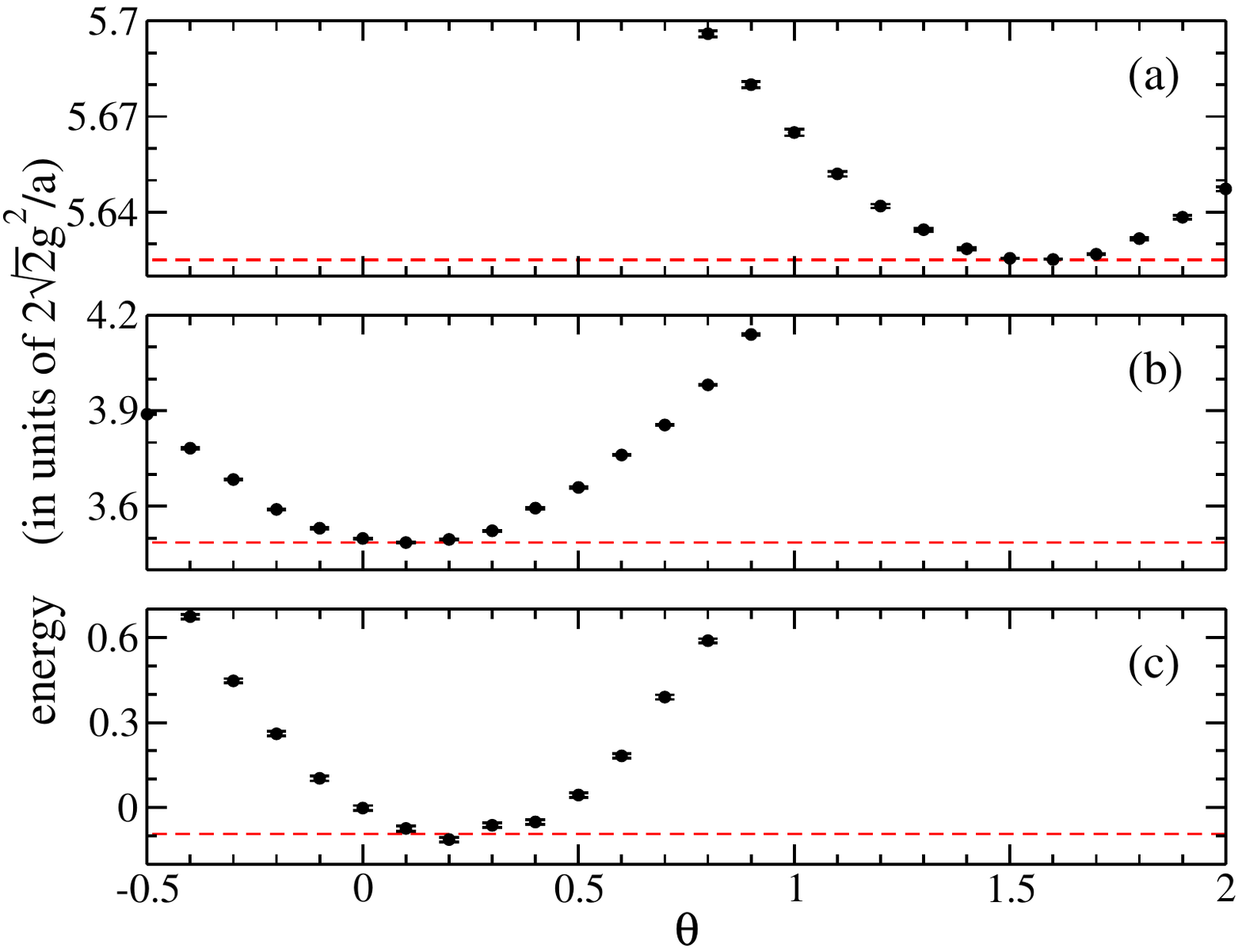}
\caption{{\bf Three energy eigenvalues for one unit cell of the triamond lattice.}
         Panels (a), (b) and (c) correspond to the third excited state,
         the first excited state, and the ground state respectively.
         The gauge coupling is $g=1$ and periodic boundary conditions are used in all directions.
         Each data point was obtained by running qiskit code with $10^4$ shots on the {\tt ibm\_qasm} noiseless simulator.
         Error bars are standard deviations.
         For comparison, red dashed lines show the true eigenvalues.\label{fig:data111}}
\end{figure}

A simple way to begin exploring this theory is to use the variational method, which will
find the lowest energy state among all states that couple to a selected ansatz.
Fig.~\ref{fig:data111} shows three energy eigenstates obtained from three different ansatze, each having just one
variational parameter.
(We name the parameter $\theta$ in every case, but they are three different parameters.)
Each of these examples is successful in finding a true eigenstate of the Hamiltonian.
The 12-qubit quantum circuits that produced these results can be found in \nameref{sec:methods}.

To gain more intuition, notice that there is no 10-sided path inside a unit cell.
This means any plaquette operator will wrap around the lattice and touch one of the gauge links twice,
thus changing that value of $j$ twice.
With our truncation to $j\in\{0,\frac{1}{2}\}$, this means only 8 gauge links are changed when a plaquette operator is
applied to the single-cell lattice.

For example, begin with $j=0$ on all 12 links (called the bare vacuum) and then apply a non-red plaquette.
The non-red plaquette touches one of the cyan links twice while avoiding the other cyan link, thus producing a state
with $j=0$ for all red and cyan links but $j=\frac{1}{2}$ for the 8 other links.
If we had applied a non-cyan plaquette instead, the result would have been exactly the same because it touches one of the red links twice.
Similarly, the non-green and non-magenta plaquettes are identical to each other ($j=0$ for green and magenta but $j=\frac{1}{2}$ for the 8 other links), and
the non-blue and non-yellow plaquettes are identical to each other ($j=0$ for blue and yellow but $j=\frac{1}{2}$ for the 8 other links).

In fact, any number of plaquette operators can be applied to the bare vacuum one after the other in any order, and
the final state will always be one of these four basis states: the bare vacuum, the non-red non-cyan plaquette,
the non-green non-magenta plaquette and the non-blue non-yellow plaquette.
The part of the Hamiltonian matrix that governs these four states is
\begin{equation}\label{eq:vacblock}
H_{\rm vac} = \frac{2\sqrt{2}g^2}{a}\left(\begin{array}{cccc} 0 & \frac{-1}{2g^4} & \frac{-1}{2g^4} & \frac{-1}{2g^4} \\
    \frac{-1}{2g^4} & 8 & \frac{-1}{32g^4} & \frac{-1}{32g^4} \\
    \frac{-1}{2g^4} & \frac{-1}{32g^4} & 8 & \frac{-1}{32g^4} \\
    \frac{-1}{2g^4} & \frac{-1}{32g^4} & \frac{-1}{32g^4} & 8 \end{array}\right) \,.
\end{equation}

Of the 4096 basis states for the periodic unit cell, only 32 obey Gauss's law.
The requirement of Gauss's law is that each site has either all three gauge links with $j=0$ or exactly one gauge link with $j=0$.
Because the triamond lattice is symmetric under color interchanges,
the 32$\times$32 Hamiltonian for Gauss-compliant states can be block diagonalized into eight 4$\times$4 blocks, including the
one shown explicitly in Eq.~(\ref{eq:vacblock}).
Each of the 32 states has exactly 0, 4, 6 or 8 of the gauge links with $j=\frac{1}{2}$.
From top to bottom, the three examples in Fig.~\ref{fig:data111} are dominated by a state where the number of gauge links with $j=\frac{1}{2}$
is 6, 4 and 0 respectively.
In particular, the bottom panel of Fig.~\ref{fig:data111} is showing the true ground state of the theory, which is built from
four basis states.

Block diagonalizing the Hamiltonian is simple enough for such a small lattice,
but phenomenological applications will require lattices with more than a single unit cell, and this is where
the Hilbert space grows rapidly.
A lattice with $N$ unit cells will have $12N$ gauge links.
Truncating each link to $j\in\{0,\frac{1}{2}\}$ leads to $2^{12N}$ basis states, and the ground state is built
from $2^{4N-2}$ of them.
Larger lattices are where the variational method, imaginary time evolution, and other algorithms implemented with error mitigation on
quantum computers have the potential to be of great value.

\subsection*{Triamond imaginary time evolution}\label{sec:triQITE}

When future computations are performed on triamond lattices, a typical first step will be to prepare the interacting vacuum.
Imaginary time evolution is a reliable method for doing so, but it is only practical if error mitigation methods are able to account
for the hardware errors.
As a first confirmation of self-mitigation for triamond lattices, consider a single unit cell with periodic boundary conditions.
The Hamiltonian matrix is Eq.~(\ref{eq:vacblock}), which can be written as
\begin{eqnarray}
\hat H_{\rm vac} &=& \frac{2\sqrt{2}g^2}{a}\bigg( 6 - 2(\hat Z_0+\hat Z_1) - 2\hat Z_1\hat Z_0 \nonumber \\
                 && - \frac{17}{64g^4}(\hat X_0 + \hat X_1 + \hat X_1\hat X_0) \nonumber \\
                 && - \frac{15}{64g^4}(\hat Z_1\hat X_0 + \hat X_1\hat Z_0 - \hat Y_1\hat Y_0) \bigg) \,.
                    \label{eq:Hvactri}
\end{eqnarray}

\begin{figure}
\includegraphics[width=85mm]{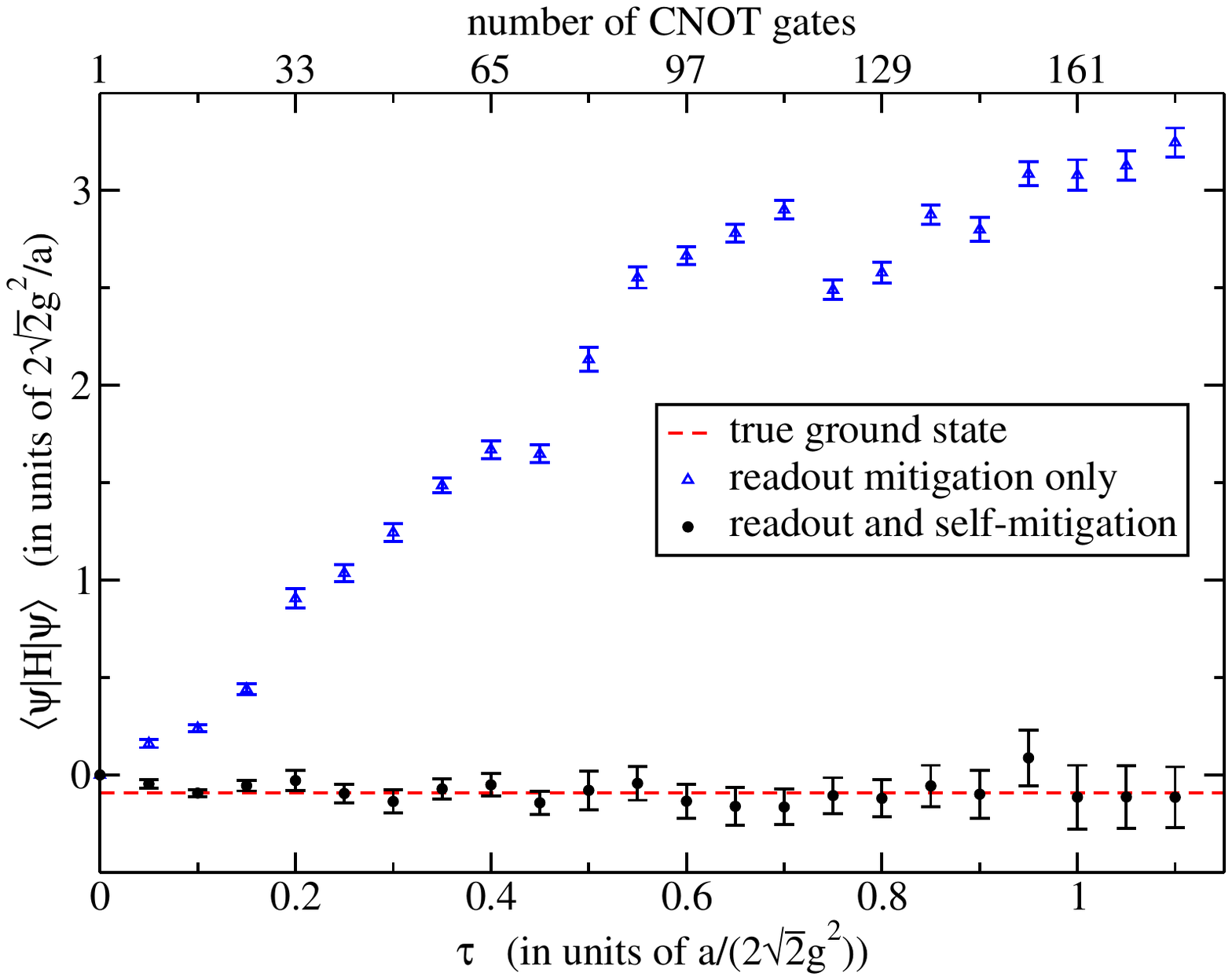}
\caption{{\bf Quantum imaginary time evolution on a triamond lattice.}
         These results are for SU(2) gauge theory with $j_{\rm max} = \frac{1}{2}$ on a triamond unit
         cell with periodic boundaries in all three spatial directions. The gauge coupling is $g=1$ and the time step is
         $\Delta\tau=0.05$ in units of $a/(2\sqrt{2}g^2)$.
         Each quantum circuit uses between 32 and 50 randomized compilings with 2000 shots per compiling.
         All quantum computations were performed on {\tt ibm\_brisbane}.
         Black data points use self-mitigation and blue data points do not.
         Error bars are 95\% confidence intervals.
         The true ground state is shown as a red dashed line.
         \label{fig:triqite2}}
\end{figure}

This is similar to the Hamiltonian for square plaquettes given in Eq.~(\ref{eq:H}), but with different coefficients and additional terms.
We adapted our code from the square plaquette study to handle the triamond case, obtaining the results displayed in
Fig.~\ref{fig:triqite2}.
Perhaps surprisingly, the unmitigated results are moving away from the true value already at the very first step of imaginary time.
Notice, however, that we are performing a particularly difficult computation: the pure noise limit is at a height of 6 (far beyond the top
of the graph) while the true result is only slightly below zero.
This means the true result contains a delicate cancellation between the large first term in Eq.~(\ref{eq:Hvactri}) and the sum of all other terms.

As expected, the unmitigated data points in Fig.~\ref{fig:triqite2} are approaching the pure noise limit as imaginary time increases.
Fluctuations in that approach are observed and were expected because the computation at each time step was usually
performed hours or days after the previous time step,
and was run on whichever qubits were performing best at that time.
In contrast, the self-mitigated results arrive at the correct value after two steps of imaginary time and remain in agreement for the
duration of our computations. 
Even the fluctuations evident in the unmitigated results are handled nicely by self-mitigation, with no visible effect
remaining in the self-mitigated data.

For $\tau<0.7$ in the units of Fig.~\ref{fig:triqite2}, we used 50 randomized compilings.  As $\tau$ increased beyond $0.7$, the number of randomized compilings was reduced linearly to just 32 randomized compilings at $\tau=1.1$.
No sensitivity to this choice is observed in the results.

\begin{table}[b]
\caption{{\bf The 16 options for randomized compiling.}
         Each of the 16 rows in this table is equivalent to a single CNOT gate on error-free hardware.
         On noisy hardware, randomly selecting from these rows is valuable for error mitigation.
         See the discussion in \nameref{sec:methods}.
        }\label{tab:rancom}
\begin{ruledtabular}
\begin{tabular}{cccc}
\multicolumn{2}{l}{Applied before the CNOT} & \multicolumn{2}{l}{Applied after the CNOT} \\
control & target & control & target \\
\hline
I & I & I & I \\
I & X & I & X \\
I & Y & Z & Y \\
I & Z & Z & Z \\
X & I & X & X \\
X & X & X & I \\
X & Y & Y & Z \\
X & Z & Y & Y \\
Y & I & Y & X \\
Y & X & Y & I \\
Y & Y & X & Z \\
Y & Z & X & Y \\
Z & I & Z & I \\
Z & X & Z & X \\
Z & Y & I & Y \\
Z & Z & I & Z
\end{tabular}
\end{ruledtabular}
\end{table}

It is also worth noticing a significant difference between the two quantum computers we have used in this work.
The square plaquette results in Fig.~\ref{fig:plot2mitb} ran on {\tt ibm\_lagos} where the CNOT gate is a native gate, but our triamond results in Fig.~\ref{fig:triqite2} ran on {\tt ibm\_brisbane} where CNOT gates are not native and are
constructed from the echoed cross-resonance (ECR) gate.
Self-mitigation has been tailored specifically to an ECR device by other authors \cite{Asaduzzaman:2023wtd} but our
code is written in terms of CNOT gates (see Table~\ref{tab:rancom}), and it is encouraging to observe that a CNOT-based self-mitigation code performs
well on the ECR device also.

\section*{Discussion}\label{sec:discussion}

In this work, SU(2) gauge theory has been truncated to $j\in\{0,\frac{1}{2}\}$ and studied on small lattices.

Self-mitigation has allowed the QITE algorithm \cite{Motta:2019yya} to find the ground state of a lattice with two square plaquettes on an IBM quantum computer.
As shown in Fig.~\ref{fig:plot2mitb}, the computation without self-mitigation did not attain the correct ground state.
The basic effect of self-mitigation is a rescaling of expectation values, where raw results from the physics circuit are divided by results from
a mitigation circuit according to Eqs.~(\ref{eq:ratio1}) and (\ref{eq:ratio2}).
The Hamiltonian's expectation value is a linear combination of the individual rescaled expectation values, leading to the large
improvement observed in Fig.~\ref{fig:plot2mitb} where, from an initial value of zero, the self-mitigated and unmitigated results ultimately move in opposite directions.
The unmitigated result is approaching the pure noise limit while the self-mitigated result finds the true value quickly and remains consistent with it.

The error bars in Fig.~\ref{fig:plot2mitb} are statistical only, and self-mitigation has clearly handled the dominant systematic error.
The mitigated result at $\tau=1.3$ happens to be furthest from the true ground state, being several standard deviations away, but subsequent time steps
agree nicely with the true ground state.
This demonstrates a valuable feature of the QITE algorithm.
Output from one time step is needed as input for the following step, and yet the algorithm recovers quickly from an outlier.

The application of QITE to a larger lattice will require more terms in the expression for $\hat A$ in Eq.~(\ref{eq:Adef}),
leading to more expectation values that need to be computed.
However, the expression for $\hat A$ will generally not require all possible Pauli strings but is instead governed by the correlation length
of the physics under consideration \cite{Motta:2019yya}.
It will be interesting to see future studies that use self-mitigated QITE to examine non-Abelian gauge theories on larger lattices.

The triamond crystal offers a systematic way to define three-dimensional lattices.
Because there are only three gauge links touching each lattice site, the quantum numbers of the links themselves are sufficient
to fully define any basis state of the lattice.
This is not true for a simple cubic lattice \cite{Muller:2023nnk}.
In fact, the triamond lattice is the only highly symmetric lattice that achieves this goal in three dimensions.
To be more precise, the triamond lattice is the only three-dimensional lattice that is strongly isotropic and has three
gauge links per site \cite{Sunada:2008}.
Strong isotropy refers to the fact that rotation (or reflection) of the lattice around any site can interchange any pair of links at that site
while leaving the physical structure of the entire lattice invariant \cite{Sunada:2008}.
In particular, a simple cubic lattice is not strongly isotropic \cite{Sunada:2008,Suizu:2022}.

Another interesting glimpse into the symmetries of the triamond lattice comes from noticing that
the six vectors comprising a triamond lattice form a regular tetrahedron.
Begin with one gauge link of each color (recall Eqs.~(\ref{eq:colorvectors1}) and (\ref{eq:colorvectors2}))
and translate each one spatially without any rotations.
The red, green and blue links will form the equilateral triangular base of the tetrahedron.
The cyan, magenta and yellow links rise from the corners of that base to meet at the top of the tetrahedron.
As might be expected from the sentences preceding Eq.~(\ref{eq:vacblock}), the cyan link of the tetrahedron touches all links
except the red one, the magenta link touches all but green, and the yellow link touches all but blue.

\begin{figure}
\includegraphics[width=81mm]{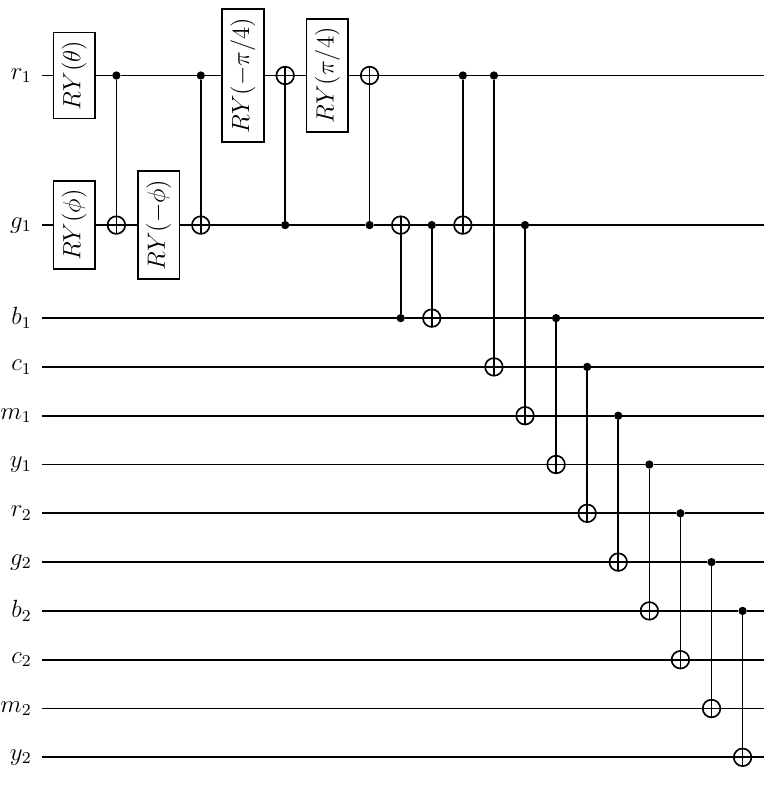}
\caption{{\bf The circuit used to obtain the ground state eigenvalue.}
         The 12 qubits are labeled by their colors, with two copies of each color appearing in the list.
         A constant angle is defined by $\phi = \arccos(1/\sqrt{3})$.
         The ground state eigenvalue for one unit cell of the triamond lattice is displayed in Fig.~\ref{fig:data111}(c).
         \label{fig:circuit2}}
\end{figure}

\begin{figure}[h]
\includegraphics[width=81mm]{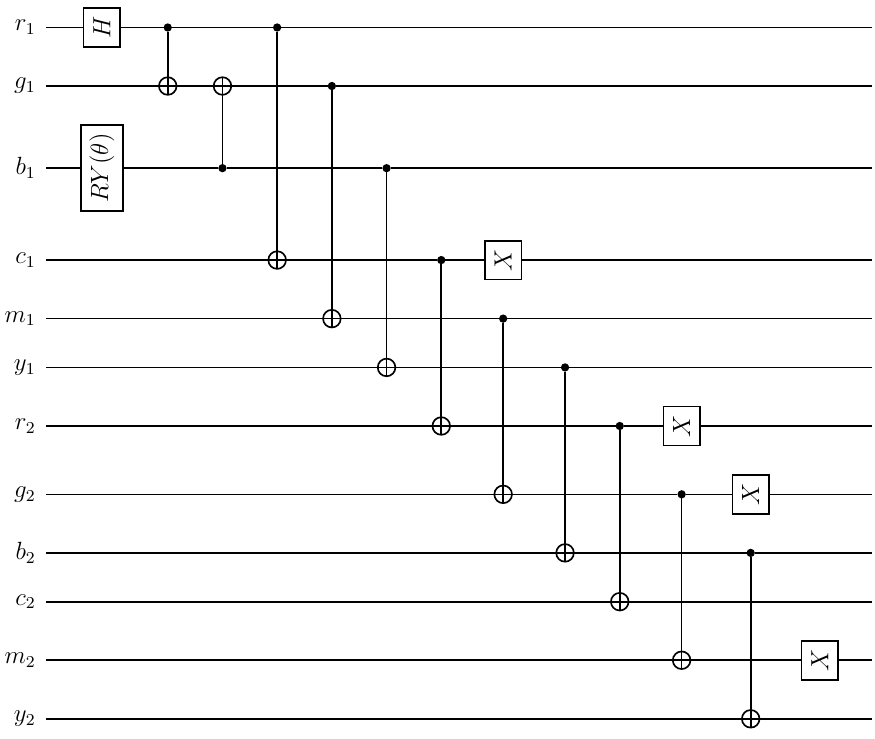}
\caption{{\bf The circuit used to obtain the first excited state eigenvalue.}
         The 12 qubits are labeled by their colors, with two copies of each color appearing in the list.
         The first excited state eigenvalue for one unit cell of the triamond lattice is displayed in Fig.~\ref{fig:data111}(b).
         \label{fig:circuit3}}
\end{figure}

For eyes used to seeing cubic lattices, the triamond might seem difficult to visualize.
However, our derivation shows that the SU(2) Hamiltonian has the familiar form of a sum over just a few plaquette
orientations, in this case six, and our quantum circuits for obtaining ground and excited states
(see Figs.~\ref{fig:circuit2} and \ref{fig:circuit3}) are quite simple due to the triamond symmetries.
The triamond lattice has only the expected parameters, namely the gauge coupling and the lattice volume, but it provides
extra gauge links inside a unit cell, representing a smaller
lattice spacing and a new trajectory for approaching the continuum limit of a gauge theory.


Fig.~\ref{fig:triqite2} shows the successful application of imaginary time evolution to a lattice in three spatial dimensions. The result is a robust
preparation of the ground state.  The triamond framework offers a systematic path toward larger lattices and relaxed truncations of
the gauge theory.  Future work in this direction can continue the progress toward a practical implementation of lattice gauge theories
on quantum computers.

\section*{Methods}\label{sec:methods}

\subsection*{Tomography for the QITE algorithm}\label{sec:tomography}

The QITE procedure \cite{Motta:2019yya} needs optimal values for the coefficients $a_{jk}$ in Eq.~(\ref{eq:Adef}).
For notational convenience the subscripts can be combined into a single subscript, so $a_{jk}\equiv a_J$ with $J$ running from 1 to 6.
The $a_J$ values can be obtained by minimizing Trotter errors, which means minimizing the difference between
\begin{equation}
\left|\Delta_0\right> \equiv \left(\frac{e^{-i\tau\hat A}-1}{\tau}\right)\left|\psi\right>
\end{equation}
and
\begin{equation}
\left|\Delta\right> \equiv -i\hat A\left|\psi\right> \,.
\end{equation}
It is more convenient to work with a scalar function rather than states, so we actually minimize
\begin{equation}
\left<\Delta_0-\Delta|\Delta_0-\Delta\right> \equiv \left<\Delta_0|\Delta_0\right> + a_Jb_J + a_JS_{JK}a_K \,,
\end{equation}
where the 6-component vector $b$ and the 6$\times$6 matrix $S$ are real-valued.
Minimization results in
\begin{equation}
a = -(S+S^T)^{-1}b \,,
\end{equation}
with
\begin{eqnarray}
b_1 &=& -i\left<\psi\right|\left[\hat H,\hat Y_0\right]\left|\psi\right>, \\
b_2 &=& -i\left<\psi\right|\left[\hat H,\hat X_1\hat Y_0\right]\left|\psi\right>, \\
b_3 &=& -i\left<\psi\right|\left[\hat H,\hat Z_1\hat Y_0\right]\left|\psi\right>, \\
b_4 &=& -i\left<\psi\right|\left[\hat H,\hat Y_1\right]\left|\psi\right>, \\
b_5 &=& -i\left<\psi\right|\left[\hat H,\hat Y_1\hat X_0\right]\left|\psi\right>, \\
b_6 &=& -i\left<\psi\right|\left[\hat H,\hat Y_1\hat Z_0\right]\left|\psi\right>,
\end{eqnarray}
and the only nonzero entries in $S+S^T$ are
\begin{eqnarray}
(S+S^T)_{11} = (S+S^T)_{22} &=& 2, \\
(S+S^T)_{33} = (S+S^T)_{44} &=& 2, \\
(S+S^T)_{55} = (S+S^T)_{66} &=& 2, \\
(S+S^T)_{12} = (S+S^T)_{21} &=& 2\left<\psi\right|\hat X_1\left|\psi\right>, \\
(S+S^T)_{13} = (S+S^T)_{31} &=& 2\left<\psi\right|\hat Z_1\left|\psi\right>, \\
(S+S^T)_{14} = (S+S^T)_{41} &=& 2\left<\psi\right|\hat Y_1\hat Y_0\left|\psi\right>, \\
(S+S^T)_{25} = (S+S^T)_{52} &=& 2\left<\psi\right|\hat Z_1\hat Z_0\left|\psi\right>, \\
(S+S^T)_{26} = (S+S^T)_{62} &=& -2\left<\psi\right|\hat Z_1\hat X_0\left|\psi\right>, \\
(S+S^T)_{35} = (S+S^T)_{53} &=& -2\left<\psi\right|\hat X_1\hat Z_0\left|\psi\right>, \\
(S+S^T)_{36} = (S+S^T)_{63} &=& 2\left<\psi\right|\hat X_1\hat X_0\left|\psi\right>, \\
(S+S^T)_{45} = (S+S^T)_{54} &=& 2\left<\psi\right|\hat X_0\left|\psi\right>, \\
(S+S^T)_{46} = (S+S^T)_{64} &=& 2\left<\psi\right|\hat Z_0\left|\psi\right>.
\end{eqnarray}
The commutators defining the elements of $b$ use the Hamiltonian from Eq.~(\ref{eq:H}) for square plaquettes
or Eq.~(\ref{eq:vacblock}) for the triamond unit cell.
The numerical computation of $b$, $S+S^T$ and $r^\prime/r$ (from Eq.~(\ref{eq:rprime})) requires nine expectation values.
Three of them can be obtained by preparing the state $\left|\psi\right>$ in a quantum computer and measuring each qubit, which gives
\begin{eqnarray}
\left<\psi\right|\hat Z_0\left|\psi\right> &=& 1 - 2P_0, \\
\left<\psi\right|\hat Z_1\left|\psi\right> &=& 1 - 2P_1, \\
\left<\psi\right|\hat Z_1Z_0\left|\psi\right> &=& 1 - 2P_{1\oplus0},
\end{eqnarray}
where $P_j$ is the probability that qubit $j$ is 1, and $P_{1\oplus0}$ is the probability that both qubits are 1.
Three other expectation values are obtained by preparing the state
$\left|\psi^\prime\right> = RY_1(-\frac{\pi}{2})RY_0(-\frac{\pi}{2})\left|\psi\right>$ in a quantum computer
and measuring each qubit, which gives
\begin{eqnarray}
\left<\psi\right|\hat X_0\left|\psi\right> &=& 1 - 2P_0^\prime, \\
\left<\psi\right|\hat X_1\left|\psi\right> &=& 1 - 2P_1^\prime, \\
\left<\psi\right|\hat X_1X_0\left|\psi\right> &=& 1 - 2P_{1\oplus0}^\prime.
\end{eqnarray}
The remaining three expectation values are obtained similarly:
Preparing $\left|\psi^{\prime\prime}\right> = RX_1(\frac{\pi}{2})RX_0(\frac{\pi}{2})\left|\psi\right>$ gives
\begin{equation}
\left<\psi\right|\hat Y_1Y_0\left|\psi\right> = 1 - 2P_{1\oplus0}^{\prime\prime}.
\end{equation}
Preparing $\left|\psi^{\prime\prime\prime}\right> = RY_0(-\frac{\pi}{2})\left|\psi\right>$ gives
\begin{equation}
\left<\psi\right|\hat Z_1X_0\left|\psi\right> = 1 - 2P_{1\oplus0}^{\prime\prime\prime}.
\end{equation}
Preparing $\left|\psi^{\prime\prime\prime\prime}\right> = RY_1(-\frac{\pi}{2})\left|\psi\right>$ gives
\begin{equation}
\left<\psi\right|\hat X_1Z_0\left|\psi\right> = 1 - 2P_{1\oplus0}^{\prime\prime\prime\prime}.
\end{equation}
Our computations create and measure these five circuits separately for each time step.
This means each one is self-mitigated, with 50 randomized compilings of its physics circuit and 50 of its mitigation circuit.

It should be noted that the QITE procedure does not require $S+S^T$ to be an invertible matrix.
For all practical purposes, $(S+S^T)^{-1}$ can be interpreted as the Moore-Penrose pseudo-inverse \cite{Moore:1920,Penrose:1955vy}
and this is what we use for our calculations.

\subsection*{Randomized compiling for CNOT gates}\label{sec:rancom}

The conversion of CNOT errors into incoherent noise is accomplished by randomizing the input to each CNOT gate in a circuit \cite{Urbanek:2021oej,ARahman:2022tkr}.
Two random Pauli gates are applied immediately before the CNOT gate, one to the control qubit and the other to the target qubit.
Specifically, each gate is chosen randomly from the set $\{I,X,Y,Z\}$.
Immediately after the CNOT gate, two Pauli gates are applied to ensure that the combined effect of Pauli gates would not change the circuit's
output on error-free hardware.
The 16 options are shown explicitly in Table~\ref{tab:rancom}.

\subsection*{Deriving the triamond SU(2) Hamiltonian}\label{sec:deriveH}

This section outlines the derivation that begins with gauge links in the form of Eq.~(\ref{eq:U}) on a triamond lattice
and arrives at the Hamiltonian in Eqs.~(\ref{eq:Htri}-\ref{eq:HB}).
The coefficients in Eqs.~(\ref{eq:Htri}-\ref{eq:HB}) are defined by their need to agree with continuum SU(2) gauge theory,
so our first step will be to expand the sum over plaquettes in powers of the lattice spacing.

Consider a general gauge link, $U(\vec w+\vec t,\hat s)$, where $\vec w$ is a white site on the lattice.
The lattice spacing $a$ enters through the vectors $\vec t$ and $\hat s$.
As a specific example, the gauge link that begins at the nearest green site and points toward the neighboring yellow site is
\begin{eqnarray}
U(\vec w+a\hat g,\frac{\hat k+\hat i}{\sqrt{2}})
&=& I + \frac{ia}{2}\left(A_x(\vec w) + A_z(\vec w)\right) \nonumber \\
& & + \frac{ia^2}{4}(\partial_x-\partial_y)\left(A_x(\vec w) + A_z(\vec w)\right) \nonumber \\
& & - \frac{a^2}{8}\left(A_x(\vec w) + A_z(\vec w)\right)^2 \nonumber \\
& & + O(a^3) \,.
\end{eqnarray}
Although not displayed here, terms at $O(a^3)$ and $O(a^4)$ must also be retained.
Performing 10 such expansions provides an expression for the non-blue plaquette, which is
\begin{eqnarray}
\mathcal{P}_{\bar b} &=& {\rm Tr}\big(U(\vec w,\hat g)U(\vec w+a\hat g,\hat y)U(\vec w+a\hat g+a\hat y,\hat m) \nonumber \\
           & & U(\vec w+a\hat g+a\hat y+a\hat m,\hat c)U^\dagger(\vec w+a\hat y+a\hat m+a\hat c,\hat g) \nonumber \\
           & & U(\vec w+a\hat m+a\hat c+a\hat y,\hat r)U^\dagger(\vec w+a\hat r+a\hat m+a\hat c,\hat y) \nonumber \\
           & & U^\dagger(\vec w+a\hat r+a\hat m,\hat c)U^\dagger(\vec w+a\hat r,\hat m)U^\dagger(\vec w,\hat r)\big) \,,
               \nonumber \\
\end{eqnarray}
and can be expanded to become
\begin{eqnarray}
\mathcal{P}_{\bar b}
&=& 2 - 2a^4{\rm Tr}\big((\partial_xA_y-\partial_yA_x+\partial_zA_y-\partial_yA_z)^2\big) \nonumber \\
& & - 4ia^4{\rm Tr}\big((\partial_xA_y-\partial_yA_x+\partial_zA_y-\partial_yA_z) \nonumber \\
& & ([A_x,A_y]+[A_z,A_y])\big) \nonumber \\
& & + 2a^4{\rm Tr}\big(([A_x,A_y]+[A_z,A_y])^2\big) + O(a^5) \,,
\end{eqnarray}
where all fields are evaluated at $\vec w$.
The expression for the non-yellow plaquette is similar to this non-blue result and summing the two of them removes
some cross terms,
\begin{eqnarray}
\mathcal{P}_{\bar b} + \mathcal{P}_{\bar y}
&=& 4 - 4a^4{\rm Tr}\big((\partial_xA_y-\partial_yA_x)^2\big) \nonumber \\
& & - 4a^4{\rm Tr}\big((\partial_zA_y-\partial_yA_z)^2\big) \nonumber \\
& & - 8ia^4{\rm Tr}\big((\partial_xA_y-\partial_yA_x)[A_x,A_y]\big) \nonumber \\
& & - 8ia^4{\rm Tr}\big((\partial_zA_y-\partial_yA_z)[A_z,A_y]\big) \nonumber \\
& & + 4a^4{\rm Tr}\big([A_x,A_y]^2+[A_z,A_y]^2\big) + O(a^5).~~~~~~
\end{eqnarray}
The sum of non-green and non-magenta plaquettes can be obtained from this by relabeling colors and $x,y,z$ subscripts.
The same is true for the sum of non-red and non-cyan.
Finally, the sum of all 6 plaquettes is
\begin{equation}
\sum_{k=1}^6\mathcal{P}_k = 12 - 8a^4{\rm Tr}\left(F_{xy}^2+F_{yz}^2+F_{zx}^2\right) + O(a^5) \,,
\end{equation}
where the field strength tensor is $F_{\mu\nu} = \partial_\mu A_\nu - \partial_\nu A_\mu + i[A_\mu,A_\nu]$.

The continuum expression for the magnetic Hamiltonian, Eq.~(\ref{eq:HBcont}), can easily be converted from an integral to
a sum,
\begin{equation}
H_B = \frac{V}{g^2}\sum_{\vec w={\rm white}}{\rm Tr}\big(F_{xy}^2(\vec w)+F_{yz}^2(\vec w)+F_{zx}^2(\vec w)\big) + O(a) \,,
\end{equation}
where $V$ is the volume per white site.
There are two white sites in each unit cell, and the volume of a unit cell is $(2\sqrt{2}a)^3$, so $V=16\sqrt{2}a^3$.
Therefore the magnetic Hamiltonian becomes
\begin{equation}
H_B = \frac{2\sqrt{2}}{g^2a}\sum_{\vec w={\rm white}}\left(12-\sum_{k=1}^6\mathcal{P}_k(\vec w)\right) + O(a) \,.
\end{equation}
The constant term has no effect on dynamics so it can be dropped and we are left with the expression given in
Eq.~(\ref{eq:HB}).

Converting the electric Hamiltonian from the continuum to the triamond lattice is straightforward.
An important detail is that instead of the volume per white site, $V$, we now need the volume per link which is $V/6$.
Begin with Eq.~(\ref{eq:HEcont}) and discretize to find
\begin{eqnarray}
H_E &=& \frac{Vg^2}{6}\sum_{n={\rm links}}{\rm Tr}\left(E_x^2(n)+E_y^2(n)+E_z^2(n)\right) + O(a) \nonumber \\
        \\
    &=& \frac{8\sqrt{2}a^3g^2}{3}\sum_{n={\rm links}}{\rm Tr}\left(E_x^2(n)+E_y^2(n)+E_z^2(n)\right) \nonumber \\
    & & + O(a)
\end{eqnarray}
in agreement with Eq.~(\ref{eq:HE}).

\subsection*{Hamiltonian and circuits for triamond unit cell}\label{sec:thetacircuits}

To obtain the results in Fig.~\ref{fig:data111}, we begin with the Hamiltonian of SU(2) gauge theory
truncated to $j\in\{0,\frac{1}{2}\}$ on a unit cell of the triamond lattice with periodic boundary conditions.
The general expression was obtained in \nameref{sec:results} and here it is written in terms of Pauli operators.

The Hamiltonian is the sum of an electric part and three magnetic parts,
\begin{equation}
H = H_E + H_{\bar r\bar c} + H_{\bar g\bar m} + H_{\bar b\bar y} \,.
\end{equation}
The electric part is a sum of contributions from all gauge links as in Eq.~(\ref{eq:ondiag}),
\begin{equation}
H_E = \frac{2\sqrt{2}g^2}{a}\sum_{k=1}^{12}\frac{1-Z_k}{2} \,.
\end{equation}
The magnetic terms correspond respectively to application of the non-red non-cyan plaquette,
\begin{eqnarray}
H_{\bar r\bar c} &=& -\frac{\sqrt{2}}{4g^2a}\prod_{k\notin\{r,c\}}X_k \nonumber \\
                 & & -\frac{3\sqrt{2}}{4g^4a}\left(\prod_{k\in\{r,c\}}\frac{1+Z_k}{2}\right)\prod_{k\notin\{r,c\}}X_k \nonumber \\
                 & & +\frac{3\sqrt{2}}{16g^4a}\left(\prod_{k\in\{r,c\}}\frac{1-Z_k}{2}\right)\prod_{k\notin\{r,c\}}X_k \,,
\end{eqnarray}
application of the non-green non-magenta plaquette,
\begin{eqnarray}
H_{\bar g\bar m} &=& -\frac{\sqrt{2}}{4g^2a}\prod_{k\notin\{g,m\}}X_k \nonumber \\
                 & & -\frac{3\sqrt{2}}{4g^4a}\left(\prod_{k\in\{g,m\}}\frac{1+Z_k}{2}\right)\prod_{k\notin\{g,m\}}X_k \nonumber \\
                 & & +\frac{3\sqrt{2}}{16g^4a}\left(\prod_{k\in\{g,m\}}\frac{1-Z_k}{2}\right)\prod_{k\notin\{g,m\}}X_k \,,~~
\end{eqnarray}
and application of the non-blue non-yellow plaquette,
\begin{eqnarray}
H_{\bar b\bar y} &=& -\frac{\sqrt{2}}{4g^2a}\prod_{k\notin\{b,y\}}X_k \nonumber \\
                 & & -\frac{3\sqrt{2}}{4g^4a}\left(\prod_{k\in\{b,y\}}\frac{1+Z_k}{2}\right)\prod_{k\notin\{b,y\}}X_k \nonumber \\
                 & & +\frac{3\sqrt{2}}{16g^4a}\left(\prod_{k\in\{b,y\}}\frac{1-Z_k}{2}\right)\prod_{k\notin\{b,y\}}X_k \,.
\end{eqnarray}
Notice that each magnetic term has 8 Pauli $X$ gates and a projector acting on the other 4 qubits.

The next thing needed to obtain the results in Fig.~\ref{fig:data111} is a quantum circuit to prepare the ansatz
containing the variational parameter.
Based on the discussion surrounding Eq.~(\ref{eq:vacblock}), we anticipate that the physical ground state will have a
contribution from the bare ground state as well as a triply degenerate contribution from three plaquette terms.
Our variational parameter $\theta$ will be used to adjust the relative strengths of these two contributions.
Our circuit is displayed in Fig.~\ref{fig:circuit2} and is responsible for the bottom panel of Fig.~\ref{fig:data111}.
This particular ansatz is quite simple, with two qubits at its core and extra CNOT gates adjusting the other qubits in
a straightforward way.
The circuit has been designed such that it requires minimal connectivity among the 12 qubits and is suitable, for example,
to run on IBM's heavy hex architecture without additional swap gates.
The center panel of Fig.~\ref{fig:data111} uses the trial state obtained from the circuit in Fig.~\ref{fig:circuit3}.
The top panel of Fig.~\ref{fig:data111} uses a trial state that differs from Fig.~\ref{fig:circuit3} simply by appending
a Pauli $X$ gate at the end of the circuit for each cyan, magenta and yellow qubit.

The final step to obtain the results in Fig.~\ref{fig:data111} is to measure the expectation value of each term in the
Hamiltonian.
Each term is a product of $X$ and $Z$ gates.
Expectation values involving only $Z$ gates can be obtained directly from measurements of the state itself.
Expectation values involving $X$ gates can be obtained by rotating into the computational basis with a
pair of Hadamard gates, $X = HZH$.
Each data point in Fig.~\ref{fig:data111} comes from the linear combination of output from 4 separate runs of the circuit,
one for $H_E$, one for $H_{\bar r\bar c}$, one for $H_{\bar g\bar m}$, and one for $H_{\bar b\bar y}$.

\section*{Data availability}

The numerical data for generating the figures are available from the authors upon request.

\section*{Code availability}

Code for calculating self-mitigated imaginary time evolution is provided at
\url{https://github.com/randylewis/SelfMitigatedQITE}.

\vspace*{4mm}

\begin{acknowledgments}
The authors thank R.~M.~Woloshyn for suggesting that self-mitigation be applied to QITE, W.~G.~Parrott for discussions about
quantum imaginary time evolution, and M.~Tsang for discussions about triamond crystals.
We thank all three for reading a draft of the manuscript.
We also acknowledge the use of IBM Quantum services for this work.
The views expressed are those of the authors, and do not reflect the official policy or position of IBM or the IBM Quantum team.
This work is affiliated with the QC4HEP Working Group and supported by Canada's Natural Sciences and Engineering Research Council.
\end{acknowledgments}

\end{document}